\documentclass{aastex61}
\received{March 7, 2017}
\revised{April 24, 2017}
\accepted{April 24, 2017}
\submitjournal{ApJL}

\begin{document}
\title{Magnetic nulls and super-radial expansion in the solar corona}

\correspondingauthor{Sarah E. Gibson}
\email{sgibson@ucar.edu}

\author[0000-0001-9831-2640]{Sarah E. Gibson}
\affil{National Center for Atmospheric Research \\
3080 Center Green Dr., 
Boulder, CO, 80301, USA}

\author[0000-0001-8929-4006]{Kevin Dalmasse}
\affil{National Center for Atmospheric Research \\
3080 Center Green Dr., 
Boulder, CO, 80301, USA}

\author{Laurel A. Rachmeler}
\affiliation{NASA Marshall Space Flight Center \\
Marshall Space Flight Center, Huntsville, AL, 35811} 

\author[0000-0002-6338-0691]{Marc L. De Rosa}
\affil{Lockheed Martin Solar and Astrophysics Laboratory\\
3251 Hanover St. B/252,  
Palo Alto, CA 94304, USA
}

\author[0000-0001-7399-3013]{Steven Tomczyk}
\affil{National Center for Atmospheric Research \\
3080 Center Green Dr., 
Boulder, CO, 80301, USA}

\author{Giuliana de Toma}
\affil{National Center for Atmospheric Research \\
3080 Center Green Dr., 
Boulder, CO, 80301, USA}

\author{Joan Burkepile}
\affil{National Center for Atmospheric Research \\
3080 Center Green Dr., 
Boulder, CO, 80301, USA}

\author{Michael Galloy}
\affil{National Center for Atmospheric Research \\
3080 Center Green Dr., 
Boulder, CO, 80301, USA}

\begin{abstract}

Magnetic fields in the sun's outer atmosphere -- the corona -- control both solar-wind acceleration and the dynamics of solar eruptions.    We present {the first clear observational evidence 
of coronal magnetic nulls in off-limb linearly polarized observations of pseudostreamers, taken by 
the Coronal Multichannel Polarimeter (CoMP) telescope}. These nulls represent regions where magnetic reconnection is likely to act as a catalyst for solar activity. CoMP linear-polarization observations also provide an independent, coronal proxy for magnetic expansion into the solar wind{, a quantity often used to parameterize and predict the solar wind speed at Earth.}
We introduce a new method for explicitly calculating expansion factors from CoMP coronal linear-polarization observations, which does not require photospheric extrapolations. We conclude that linearly-polarized light is a powerful new diagnostic of critical coronal magnetic topologies and the expanding magnetic flux tubes that channel the solar wind.

\end{abstract}


\section{Background} \label{sec:intro}

Measuring
the strength and direction of the solar coronal magnetic field is a fundamental requirement for understanding solar activity and evolution, but remains a significant challenge. When simplifying assumptions are used such as the current-free or potential limit to the magnetic field in the corona, observations of the magnetic field at the solar 
surface
may 
be used to approximate the coronal field.
Such photospheric extrapolations -- e.g., the potential-field-source-surface (PFSS) model -- provide a good first-order characterization of the global coronal and interplanetary magnetic field \citep{newkirk_69,schatten_69}.  However, 
this model ignores currents, removing
the component of the magnetic energy that drives solar eruptions.  Another possibility is to use the full vector of the magnetic field at the photospheric boundary and solve a non-linear force-free field (NLFFF) problem (under the assumption that magnetic forces dominate in the highly-conductive corona). However, issues such as a lack of magnetic dominance at the photosphere 
lead
to inconsistencies and non-uniqueness of NLFFF models \citep{Derosa09}.  The observed photospheric boundary on its own is therefore not sufficient to model the coronal magnetic field, motivating the explicit incorporation of coronal observations into magnetic models.

The Coronal Multichannel Polarimeter (CoMP) \citep{tomczyk_08} obtains daily observations of coronal emission-line polarization in the solar atmosphere, providing
unique constraints on 
coronal magnetic models. Polarimetry has been used to determine magnetic fields at the solar surface since Hale's observations in 1908. More recently, routine measurements of magnetism in its lower atmosphere, or chromosphere, have become possible \citep{lopeariste_02,harvey_06}. At higher coronal altitudes, emission-line measurements of circular polarization -- sensitive to line-of-sight (LOS) {oriented} magnetic-field strength -- have been rare \citep{harve_69,lin_00,lin_04,tomczyk_08}. Obtaining routine circular-polarization measurements at the level of one Gauss in 15 minutes generally requires a larger-aperture telescope than the 20-cm CoMP \citep{tomczyk_16}. On the other hand, linear polarization, which is approximately 100 times brighter than circular polarization,
is measured by CoMP on a daily basis. As we now demonstrate, linear-polarization observations are ideally suited 
for probing
certain topological and expansion properties of the coronal magnetic field.

\section{Linear-polarization diagnostics: magnitude and Azimuth}\label{sec:lindiag}

The Fe XIII coronal emission line at $1074.7~nm$ is linearly polarized when unpolarized light emitted from lower layers of the solar atmosphere  undergoes scattering in the corona.
However, magnetic fields in the corona partially depolarize it, a process known as the Hanle effect \citep{sahalbrechot_77, querfeld_84,arnaud_87}.
In the corona, at this wavelength, the depolarization 
of the scattered light 
preserves information about
the direction of the magnetic field, but not its strength. 
In particular, if the local magnetic vector has an angle $\vartheta_B$  relative to the solar radius vector ($\hat{r}$) that is less than a critical ``van Vleck'' angle \citep{vvleck_25}, its projection into the plane of sky (POS) lies parallel to the observed linear-polarization vector.  However, if $\vartheta_B$ is greater than this critical angle,  then the direction of the linear-polarization vector (known as the {\it azimuth}) is rotated $90^\circ$.
The critical van-Vleck angle, $\vartheta_B = 54.74^\circ$, arises because of an atomic-alignment dependence of the form $3 \cos^2 \vartheta_B = 1$; at this critical angle the fraction of linearly-polarized light (linear-polarization {\it magnitude}) is zero.

\begin{figure}[ht!]
\begin{center}
\includegraphics
[trim={2.2in 2.2in 0.5in 2.2in},clip,width=6in]
{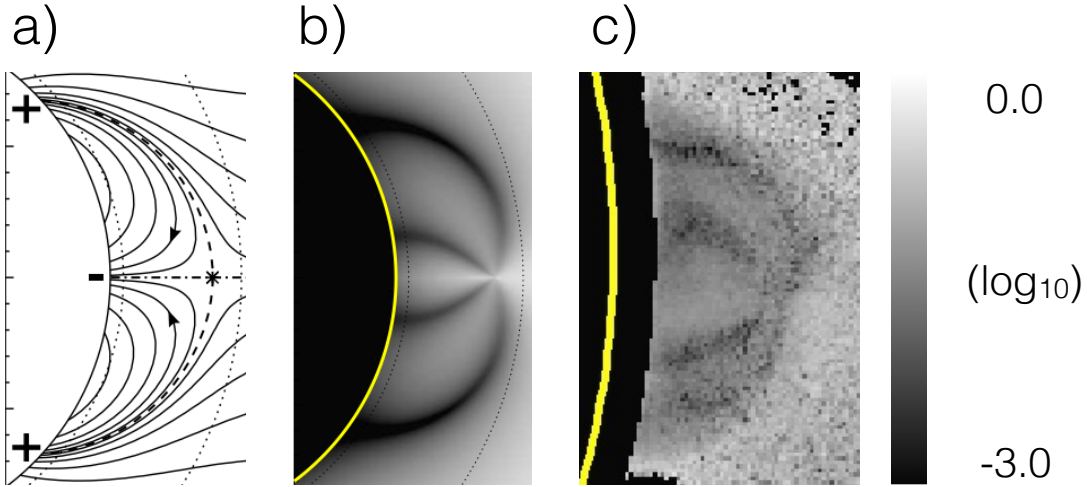}
\caption{{\bf Coronal pseudostreamer magnetic topology and associated magnetic null produces 
a characteristic signature in linear-polarization magnitude.} a-b) represent a prediction of this signature made by \citet{rachmeler_14}, and c) their manifestation in CoMP observations.  a) Magnetic field lines are illustrated within a pseudostreamer, showing two closed-field loops surrounded by unipolar open field with the central magnetic null {(i.e., where the magnetic field strength equals zero)} indicated by an asterisk.  b) Linear-polarization magnitude fraction ($L/I = \sqrt{(Q^2+U^2)}/I$, where $I, Q, U$ are Stokes vector components) is synthesized for this magnetic topology, resulting in a predicted polarization signature of pseudostreamer 
topology. 
Dark features outline three linear-polarization lobes and indicate van-Vleck angle crossings (see text); the magnetic null in (a) coincides with the intersection of these lobes.  c) CoMP linear-polarization ($L/I$) observations of a South-pole pseudostreamer on Nov. 26, 2014 demonstrate
for the first time that a magnetic null or nulls may be 
identified
with coronal polarimetry.  {The solar photosphere is indicated by the yellow curves in (b) and (c).}
}
\end{center}
\label{pseudonull}
\end{figure}

CoMP observations and forward modeling have demonstrated that,
 {\it in certain magnetic topologies},
the 
loci of van-Vleck angle crossings can be picked out of linear-polarization magnitude observations as coherent
and elongated dark features.
The corona is optically thin in the Fe XIII line, and observations are intensity-weighted along the LOS{, so multiple bright structures of varying magnetic orientations can in general contribute to the linear polarization observed by CoMP.  Thus,} the presence of distinct   
van-Vleck loci 
implies a degree of isolation and/or extension along the LOS of the magnetic structure. Under such conditions, these loci have proved useful for identifying coronal topologies.
For example, the linear-polarization signatures expected from forward modeling cylindrical vs. spheromak magnetic flux ropes are distinct \citep{rachmeler_13}. CoMP observations have provided one example of a possible spheromak \citep{dove_11};
far more examples have been found of coronal prominence cavities \citep{gibson_15} that, for many sizes and shapes, possess ``lagomorphic'' (rabbit-headed) signatures in linear-polarization magnitude, consistent with a cylindrical magnetic flux-rope topology \citep{ula_13,ula_14}. 

\section{Coronal magnetic nulls from linear-polarization magnitude}\label{sec:magtop}

{Coronal pseudostreamers and associated magnetic nulls} are likely locations for a restructuring of the magnetic field to be triggered, potentially leading to coronal mass ejections (CMEs) \citep{antio_break, torok_11, lynch_13}. {Pseudostreamers
consist of multipolar closed field} surrounded by unipolar open magnetic field 
\citep{hundhausen_solarwind, zhaowebb_03, wang_07}. They are distinguished from so-called ``helmet'' (or bipolar) streamers, which are surrounded by open field of 
opposite
polarities.  {Magnetic null points are expected in pseudostreamer topologies, as shown in Fig. 1a. Although pseudostreamers are three-dimensional magnetic structures, this two-dimensional idealization is a reasonable representation of the middle portion of a pseudostreamer that is oriented along an observer's LOS.  As discussed above and further demonstrated below, such an optimal orientation is in fact a necessary selection factor for obtaining clear linear polarization signatures. 
We will refer to magnetic nulls throughout this paper, however, we note that this effectively two-dimensional interpretation does not distinguish between a true, three-dimensional magnetic null point 
and topologically related structures such as a
superposition of a line of nulls, a magnetic separator, or a hyperbolic flux tube (see, e.g., \citet{Titov02,aulanier_05,parnell_10,Titov11}).}

Forward models of coronal pseudostreamer magnetic configurations have predicted a characteristic signature of three linear-polarization lobes (bright regions outlined by van-Vleck dark features; see Fig. 1b), which come to a point at the magnetic null \citep{rachmeler_14}.  In that work, CoMP data were presented which were consistent with the predicted pseudostreamer configuration, but which did not extend high enough to include the magnetic null.  We have now observed a pseudostreamer with CoMP data that unequivocally matches the forward-modeled prediction of the pseudostreamer topology, and explicitly contains the signature of the converging lobes indicative of magnetic null(s) at their top
(Fig. 1c).

\begin{figure}[ht!]
\begin{center}
\includegraphics[trim={.55in .87in .05in .87in},clip,width=7.3in]{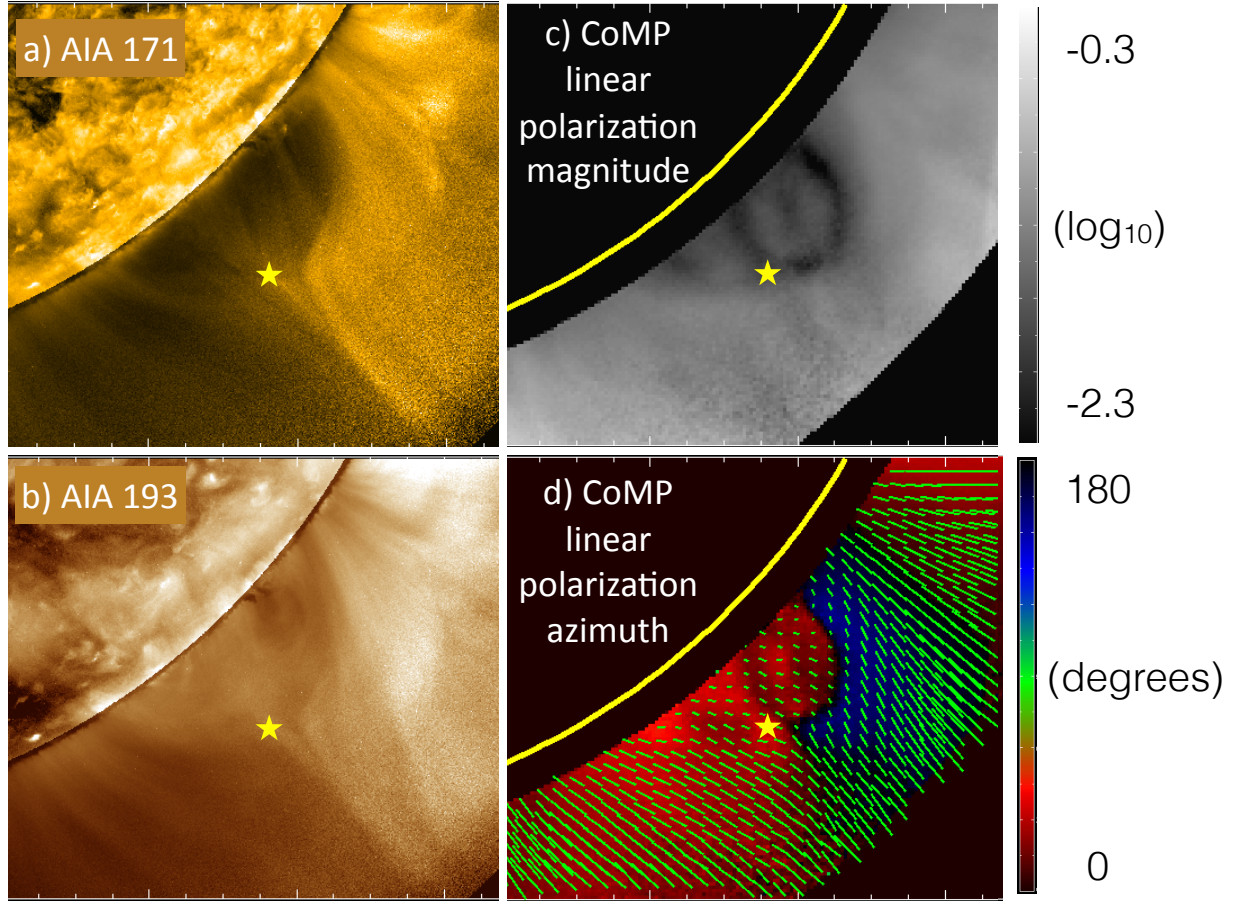}
\caption{
{\bf Observations of a pseudostreamer with a magnetic null and non-radial magnetic expansion.
}
a) AIA 171 \AA~and b) AIA 193 \AA~
intensity (19:30 UT, April 18, 2015). 
c) CoMP linear-polarization magnitude ($L/I$)
and d) azimuth = $0.5 \tan{U/Q}$ (150-image average; 18:43-19:58 UT). Azimuth direction is indicated by green vectors and by color table 
(black=radial; blue=clockwise tilt; red=counterclockwise tilt). 
The yellow stars mark the intersection of the CoMP linear-polarization magnitude lobes determined {by eye} from c). {The solar photosphere is indicated by the yellow curves in (c) and (d).}
}
\label{pseudoexp1}
\end{center}
\end{figure}

\begin{figure}[ht!]
\begin{center}
\includegraphics[trim={.55in .87in .05in .87in},clip,width=7.3in]{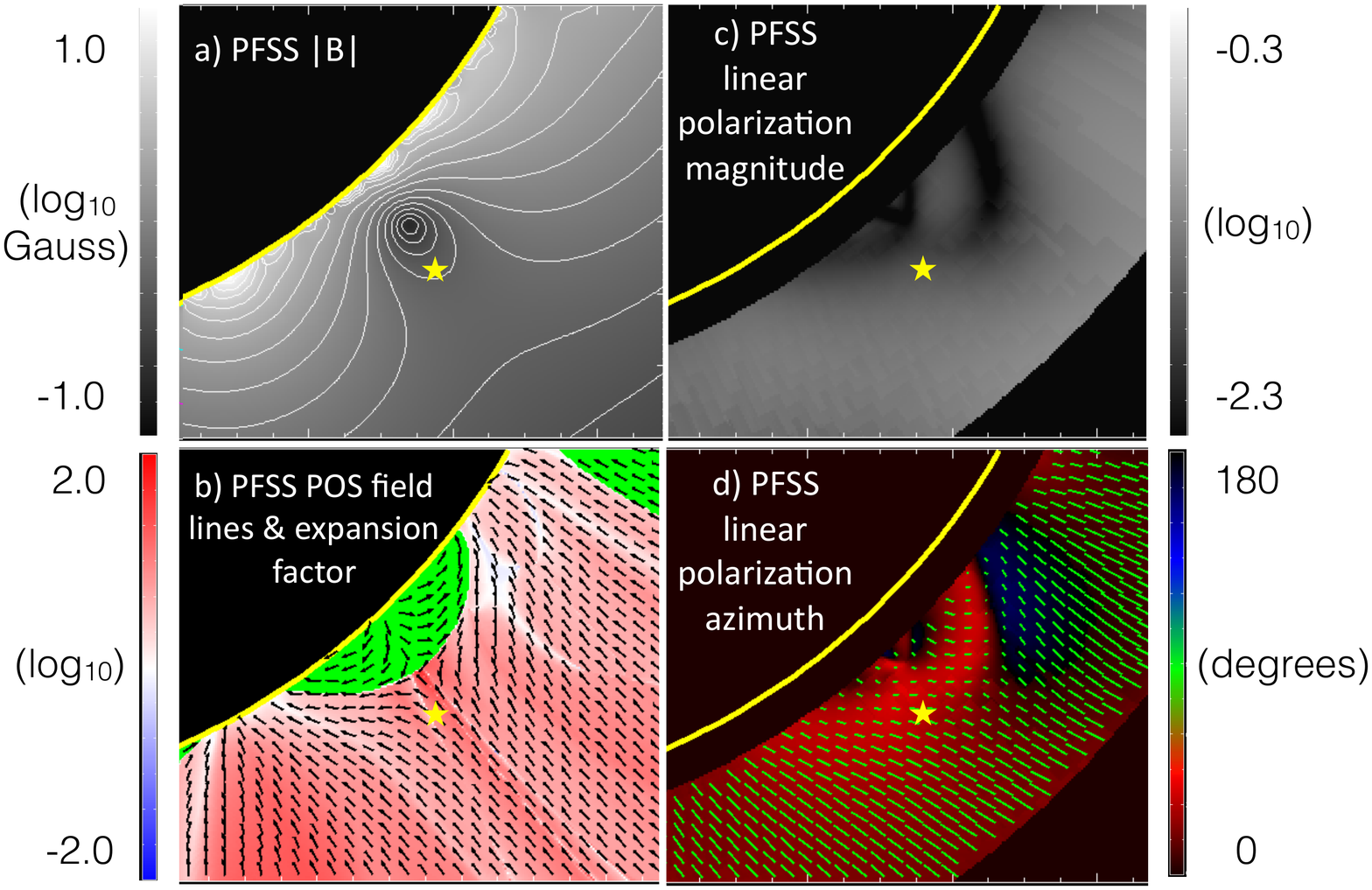}
\caption{
{\bf 
Pseudostreamer magnetic null height and non-radial magnetic expansion are underestimated by the potential-field-source-surface (PFSS) model.
}
a) PFSS magnetic field magnitude at 19:30 UT in the plane of the sky (POS).
b) PFSS magnetic expansion factor $f$ (see Appendix, Equation \ref{Eq-ExpFac-L})
 (radial expansion = white, super-radial expansion = red, sub-radial expansion = blue;
closed fields = green, POS-projected magnetic vectors = black). c,d) PFSS synthetic linear-polarization magnitude and azimuth as in Fig. 2. {Note that the synthetic (forward-modeled) green azimuth vectors shown in d) can be compared to the POS black magnetic vectors in b) with two important caveats: the azimuth is a LOS weighted integral, and the azimuth vectors rotate $90^\circ$ when the local magnetic field  is oriented at an angle greater than $54.74^\circ$ relative to the radial direction. As discussed in the text, symmetry along the LOS and local field orientation in the upper-right region of d) make the azimuth vectors in this region (shown in Fig. 4) representative of an effectively two-dimensional POS magnetic field.} The yellow stars mark the intersection of the CoMP linear-polarization magnitude lobes determined {by eye} from Fig. 2c). {The solar photosphere is indicated by the yellow curves.}
}
\label{pseudoexp2}
\end{center}
\end{figure}

Figs. 2 and 3 present another example of a pseudostreamer, and directly compare CoMP data to  PFSS model predictions.  Synthetic data are generated using the `\texttt{pfss}' and  `\texttt{FORWARD}' \citep{gibson_16} packages of SolarSoft IDL \citep{freeland_98}. Plasma weighting along the LOS is achieved by defining density radial profiles consistent with hydrostatic equilibrium: closed-field regions are assigned profiles that match coronal streamer observations \citep{gibstreamwsm}, while open-field regions are given profiles to match coronal hole observations \citep{guhawsmhole}. 

Note that, although more asymmetric than the pseudostreamer of Fig. 1, the lobes of this pseudostreamer are apparent in linear polarization for both CoMP (Fig. 2c) and forward-modeled PFSS (Fig. 3c). PFSS magnetic field plots
show this topology even more clearly (Figs. 3a-b). However, a visual inspection of the CoMP observations indicate a higher altitude null than the PFSS model possesses. The location of this CoMP null coincides with the base of the sharp linear feature extending above the pseudostreamer dome {and up into the pseudostreamer stalk, as seen by} the Solar Dynamics Observatory Atmospheric Imaging Assembly (AIA) images (Figs. 2a-b). {We note that this feature is somewhat to the left of the center of the dome, but also that the CoMP linear polarization feature is similarly asymmetric.}. The PFSS model also does not fully capture the bulging nature of the northward lobe seen in the CoMP images. The AIA 193~{\AA} image (Fig. 2b) shows the presence of a prominence cavity within this lobe, so, it is possible the expanded lobe is a consequence of currents associated with a flux rope that are missed by the (current-free) PFSS model. 

\section{Coronal nonradial magnetic expansion from linear-polarization Azimuth}\label{sec:expfac}

Another important coronal property is the height-dependent expansion of open magnetic field surrounding closed-field structures such as helmet streamers and pseudostreamers. The degree of super-radial expansion of open-field regions is of particular interest because it has been shown to be inversely correlated to solar-wind speed \citep{wangsheeley_90}. A magnetic expansion factor, 
$f$ 
(see 
Appendix, Equation \ref{Eq-ExpFac-L}; also 
Fig. 3b), may be calculated from measurements at the solar photosphere 
along with, e.g., PFSS model evaluations at higher altitudes. 
This is then a standard input into empirical models that predict solar-wind speed at the Earth {(e.g.,} \citet{arge_00}). 

CoMP linear-polarization observations present us with a new resource for probing magnetic flux-tube expansion as a function of height in the low corona, independent of model extrapolations. 
Figs. 2d) and 3d)
illustrate that van-Vleck crossings manifest not only as dark features in linear-polarization magnitude, but also as discontinuities in linear-polarization azimuth, so that diverging vs. converging magnetic fields have distinct appearances. 
Thus, azimuths crossing the van-Vleck angle -- for example within the closed-field loops that make up the pseudostreamer-- result in sharp jumps from red to blue.  In contrast, the diverging magnetic fields to the north -- associated with magnetic flux tubes open to the solar wind -- have a gradual transition from blue to red and no van-Vleck crossing. {However, the sharp jump at the van-Vleck angle may be blurred if, for example, the locus of these angles do not superpose along the LOS; this appears to be the case for the southern van-Vleck crossings 
for both the CoMP observations and the synthetic PFSS data.}

\begin{figure}[ht!]
\begin{center}
\includegraphics[trim={.1in .5in .1in .5in},clip,width=6in]{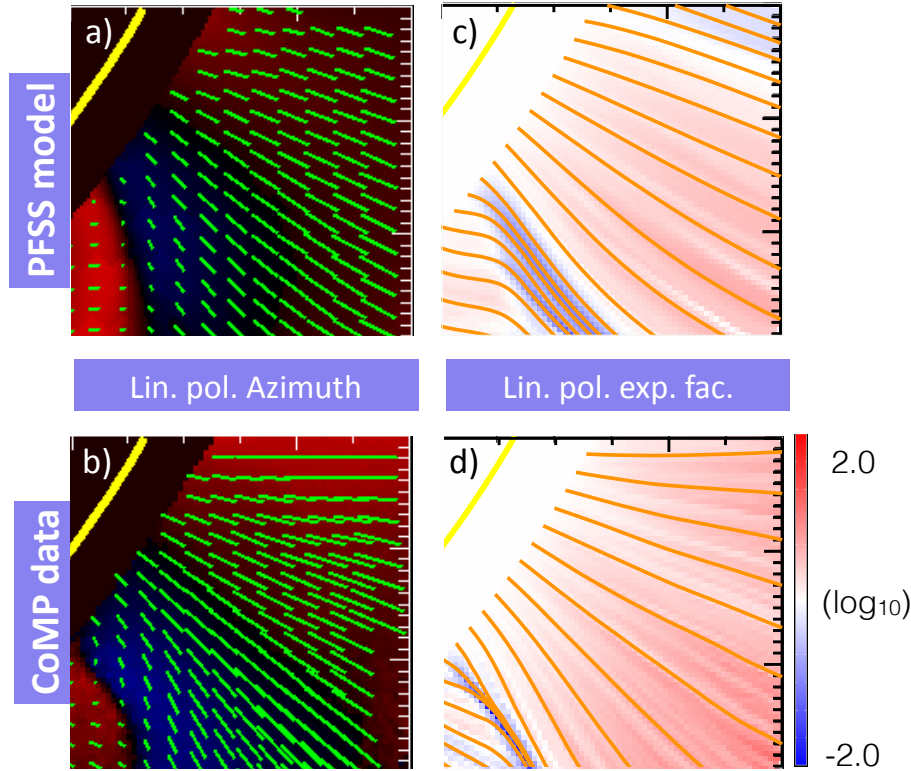}
\caption{{\bf Non-radial expansion around pseudostreamer is quantified as linear-polarization expansion factor ($LPF$) in observational COMP data and
synthetic PFSS data.}
These figures show a zoom-in of the upper-right region of Figs. 2 and 3.
Azimuths are from a) PFSS and from b) CoMP 
(the van-Vleck inversion is seen in the bottom left).
Corresponding integrated polarization field lines are shown in c) and d) 
overplotted on the $LPF$ calculated from them (radial expansion = white, super-radial expansion = red, sub-radial expansion = blue) (see Appendix). {The solar photosphere is indicated by the yellow curves.
} }
\end{center}
\label{expcalc}
\end{figure}

A visual comparison of the CoMP-observed azimuth angle (Fig. 2d) and the PFSS {forward-modeled} azimuth angle (Fig. 3d) suggests that the PFSS {extrapolation} underestimates the super-radiality of magnetic expansion to the north of the pseudostreamer.
To further investigate this, we
 create maps of an expansion-factor proxy from both the CoMP and the synthetic PFSS data
 (Fig. 4). Using a predictor-corrector Euler algorithm 
and
bilinear interpolation, we integrate polarization field lines following the azimuth down to the lower CoMP occulter.  
A linear-polarization expansion factor ($LPF$) is then determined from the local polarization
field-line deviation (see Appendix). We find that the CoMP data in this region indicates a more super-radial expansion than the PFSS model predicts. {We note that this technique relies on an interpretation of the azimuth as an effectively two-dimensional POS field which only holds true in regions that are not heavily impacted by LOS effects, and that have magnetic field orientation tilted less than the van-Vleck angle from radial.  The clear identification and localization of the van-Vleck inversion to the bottom-left  (see also Fig. 3) gives us confidence that both conditions are met for the diverging, open magnetic field shown in this figure and used to calculate the $LPF$.}

\section{Discussion} \label{sec:disc}

We have shown how CoMP linear-polarization observations may be used to quantify both magnetic topology (height of magnetic nulls) 
and super-radial expansion ($LPF$ as a function of height and latitude). The detection of magnetic nulls provides a new way to identify critical points in the coronal field that facilitate solar eruptions.
The linear-polarization expansion factor proxy, $LPF$, provides a new means for validating PFSS (and other) model predictions.

It has been pointed out that pseudostreamer-sourced solar wind may be inconsistent with the empirical relationship between PFSS-model-derived expansion factor and wind speed \citep{riley_11}. In particular, the authors observed solar wind associated with a pseudostreamer to be slow, in contrast to the fast wind predicted by low PFSS-derived expansion factors. 
{\citet{wang_12} have also pointed out that the null-point height affects magnetic expansion, and that nonmonotonically expanding flux tubes may result in an overestimate of solar wind speed.} 
The CoMP observations we have presented demonstrate how discrepancies {between observations and the empirically-predicted solar wind speed} could at least partially arise from an underestimate of magnetic expansion {and null-point height} by the PFSS model due to intrinsic limitations 
of 
photospheric extrapolations. {It is also possible that reconnections at the open-closed magnetic boundary rather than flux-tube expansion is the dominant physical process controlling the solar wind speed associated with pseudostreamers, as has been argued in other models (e.g., \citet{antiochos_11,delzanna_11}). These models too require accurate determination of coronal magnetic topology, in particular magnetic null point location.}
In conclusion, coronal linear-polarization measurements represent a valuable new type of data for {validating and developing} 
next-generation 
coronal and solar-wind acceleration models.


Acknowledgements. NCAR is supported by the National Science Foundation. SEG, KD, GdT and ST acknowledge support from the Air Force Office of Space Research, FA9550-15-1-0030. KD acknowledges funding from the Computational and Information Systems Laboratory, the High Altitude Observatory, and the Advanced Study Program. AIA data are courtesy of NASA/SDO and the AIA, EVE, and HMI science teams. We acknowledge helpful discussions with Anna Malanushenko, Ed Deluca, Antonia Savcheva and Scott McIntosh.

\appendix \section{Computing a Linear-Polarization Expansion Factor ($LPF$)}

\subsection{Background} 
A magnetic expansion factor measures the degree to which the cross sections of open magnetic flux tubes expand (or contract) in the solar corona as compared with their cross sections at the solar photosphere. In particular, magnetic expansion factor can be defined as:
	\begin{eqnarray}
		\label{Eq-ExpFac-GM1}
		f (\vec{r}, \Phi) & = & \frac{\mathcal{S} (\vec{r}, \Phi)}{\mathcal{S} (\vec{R}_{\odot}, \Phi)} \cdot \frac{R_{\odot}^2}{r^2}   \,,      
	\end{eqnarray}
where $\mathcal{S} (\vec{r}, \Phi)$ is the cross-sectional area of a tube carrying magnetic flux, $\Phi$, at position, $\vec{r}$, along its axis, $R_{\odot}$ is the solar radius, and $R_{\odot}^2/r^2$ is a correction factor accounting for the natural expansion of surfaces as $r$ increases in spherical geometry (referred to as ``radial expansion'' in the text). 

From Eq.~\ref{Eq-ExpFac-GM1}
it is possible to define a completely local expansion factor
which does not depend on the surface cross section, $\mathcal{S}$. For this we employ ``elemental'' flux tubes, which possess an infinitesimal cross section surrounding a magnetic field line. Let $\mathrm{d} \Phi = B \mathrm{d} \mathcal{S}$ be an elemental magnetic flux tube, where $B$ is the magnetic field strength along the flux tube axis and $\mathrm{d} \mathcal{S}$ is the elemental flux tube cross section. Then Eq.~\ref{Eq-ExpFac-GM1} 
applied to elemental magnetic flux tubes leads to
	\begin{eqnarray}
		\label{Eq-ExpFac-LM1}
		f (\vec{r}, \mathrm{d} \Phi) & = & \frac{\mathrm{d} \mathcal{S} (\vec{r}, \mathrm{d} \Phi)}{\mathrm{d} \mathcal{S} (\vec{R}_{\odot}, \mathrm{d} \Phi)} \cdot \frac{R_{\odot}^2}{r^2} =  \frac{\left( \mathrm{d} \Phi / B (\vec{r}) \right)}{\left( \mathrm{d} \Phi / B (\vec{R}_{\odot}) \right)} \cdot \frac{R_{\odot}^2}{r^2}    \,,   \nonumber   \\
		\label{Eq-ExpFac-L}
		 & = & \frac{B (\vec{R}_{\odot})}{B (\vec{r})} \cdot \frac{R_{\odot}^2}{r^2} = f(\vec{r})   \,, 
	\end{eqnarray}
where $f(\vec{r})$ is the local expression of the expansion factor.

When the magnetic field is known in the entire volume, one can compute the local  expansion factor from Eq.~\ref{Eq-ExpFac-L} 
by simply integrating field lines and evaluating the magnetic field strength at each step along the field lines \citep{wangsheeley_90}.  When the magnetic field direction is known, but not the magnetic field strength, we must return to Eq.~\ref{Eq-ExpFac-GM1}, 
which we will use to define and justify a linear-polarization expansion factor proxy, $LPF$, determined, e.g., from CoMP linear polarization measurements.


\subsection{$LPF$ proxies} 

To evaluate the $LPF$, 
we build upon methodology developed to compute the squashing factor of magnetic flux tubes \citep{Pariat12,tassev_sub}. The squashing factor is a measure of the distortion of magnetic flux tubes that provides a means of identifying quasi-separatrix layers (QSLs) in a 3D magnetic field \citep{demoulin_96,Titov02}, and has been extensively used in solar flare analysis to relate photospheric flare ribbons to the topology of the coronal magnetic field \citep{schmieder_97,masson_09,Savcheva12a,dalmasse_15}. QSLs are regions of strong gradients in connectivity of magnetic field lines, and are likely regions for current sheet/layers formation and magnetic reconnection \citep{aulanier_05,janvier_13}.

{The method is based on integrating a field line and calculating the field line deviation along it, $\delta \vec{r}$, from the local variations of the magnetic field vector as obtained from its local interpolation \citep{longcope_94,titov_03,tassev_sub}.} 
Once $\delta \vec{r}$ has been integrated along a field line, we can compute $\delta \vec{r}_{\perp}$, the component of $\delta \vec{r}$ perpendicular to the field line. For a 3D flux tube of circular cross section, $\delta \vec{r}_{\perp} (s)$ effectively represents the diameter of the flux tube at position $\vec{r}_s$ along the flux tube axis. $\delta \vec{r}_{\perp} (s)$ can thus be used to compute an estimate of the local expansion factor using
Eq.~\ref{Eq-ExpFac-GM1}.

Because of the LOS integration effect, the azimuth computed from CoMP only provides an {effectively} two-dimensional magnetic field direction. Thus, the expansion factor computed from CoMP linear-polarization measurements can only be a proxy of the true, local expansion factor of the coronal magnetic field. Using Eq.~\ref{Eq-ExpFac-GM1}, 
we define three, slightly different, proxies of expansion factor
	\begin{eqnarray}
		\label{Eq-Expfac-Proxy1}
		f^{\mathrm{(3D)}} ( \vec{r} (s) ) & = & \left( \frac{\| \delta \vec{r}_{\perp} (s) \|}{\| \delta \vec{r}_{\perp} (s_{\odot}) \|} \right)^2 \cdot \frac{R_{\odot}^2}{r^2}  \,,  \\
		\nonumber \\
		\nonumber \\
		\label{Eq-Expfac-Proxy2}
		f^{\mathrm{(2.5D, sph.)}} ( \vec{r} (s) ) & = & \frac{\| \delta \vec{r}_{\perp} (s) \| \sin \left( \theta (s) \right) }{\| \delta \vec{r}_{\perp} (s_{\odot}) \| \sin \left( \theta (s_{\odot}) \right) } \cdot \frac{R_{\odot}}{r}  \,,  \\
		\nonumber \\
		\nonumber \\
		\label{Eq-Expfac-Proxy3}
		f^{\mathrm{(2.5D, cyl.)}} ( \vec{r} (s) ) & = & \frac{\| \delta \vec{r}_{\perp} (s) \|}{\| \delta \vec{r}_{\perp} (s_{\odot}) \|} \cdot \frac{R_{\odot}}{r}  \,.
	\end{eqnarray}
Eq.~\ref{Eq-Expfac-Proxy1}
assumes that the CoMP data is associated with 3D magnetic flux 
tubes of circular cross section, 
$\mathcal{S} (\vec{r} (s)) = \pi (\| \delta \vec{r}_{\perp} (s) \| /2)^2$, and that axes are 
in the POS. 
Eq.~\ref{Eq-Expfac-Proxy2} 
assumes axisymmetry around the North-South 
axis of the spherical Sun, in which case the cross section of a flux tube is 
$\mathcal{S} (\vec{r} (s)) = \| \delta \vec{r}_{\perp} (s) \| 2 \pi r(s) \sin \left( \theta (s) \right)$. 
Finally, Eq.~\ref{Eq-Expfac-Proxy3}
assumes invariance in the direction of the line of 
sight, meaning that the Sun appears as a cylinder of infinite length, $L$, in the LOS direction; 
magnetic flux tubes thus have a cross section, 
$\mathcal{S} (\vec{r} (s)) = \| \delta \vec{r}_{\perp} (s) \| L$, and the correction factor 
that must be applied to remove the natural expansion of surfaces as $r$ increases is now 
$R_{\odot}/r$. For the case presented in this paper, we have compared all three $LPF$ proxies and found that although they vary somewhat in magnitude, they are very similar in structure.  The conclusions we make comparing the CoMP and PFSS $LPF$ measures are independent of which measure is used; in Figure~4 we show the 3D proxy defined by Eq.~\ref{Eq-Expfac-Proxy1}.

\end{document}